# BEMEval-Doc2Schema: Benchmarking Large Language Models for Structured Data Extraction in Building Energy Modeling

Yiyuan Jia[1*], Xiaoqin Fu[2], Liang Zhang[2,3]

1. Big Ladder Software, Denver, Colorado, USA
2. University of Arizona, Tucson, Arizona, USA
3. National Laboratory of the Rocikes, Golden, Colorado, USA
*Corresponding Author: yiyuan.jia@bigladdersoftware.com

## Abstract

Recent advances in foundation models, including large language models (LLMs), have created new opportunities to automate building energy modeling (BEM). However, systematic evaluation has remained challenging due to the absence of publicly available, task-specific datasets and standardized performance metrics. We present BEMEval, a benchmark framework designed to assess foundation models' performance across BEM tasks. The first benchmark in this suite, BEMEval-Doc2Schema, focuses on structured data extraction from building documentation, a foundational step toward automated BEM processes.

BEMEval-Doc2Schema introduces the Key–Value Overlap Rate (KVOR), a metric that quantifies the alignment between LLM-generated structured outputs and ground-truth schema references. Using this framework, we evaluate two leading models (GPT-5 and Gemini 2.5) under zero-shot and few-shot prompting strategies across three datasets: HERS L100, NREL iUnit, and NIST NZERTF. Results show that Gemini 2.5 consistently outperforms GPT-5, and that few-shot prompts improve accuracy for both models. Performance also varies by schema: the EPC schema yields significantly higher KVOR scores than HPXML, reflecting its simpler and reduced hierarchical depth. By combining curated datasets, reproducible metrics, and cross-model comparisons, BEMEval-Doc2Schema establishes the first community-driven benchmark for evaluating LLMs in performing building energy modeling tasks, laying the groundwork for future research on AI-assisted BEM workflows.

## Introduction

Buildings account for roughly 30% of global energy consumption and 26% of carbon dioxide emissions (IEA, 2023), making them a central focus in the worldwide effort to achieve decarbonization and mitigate greenhouse gas emissions (Xiang, Ma et al., 2022). Building energy modeling (BEM) serves as a pivotal tool in this endeavor, enabling the analysis and optimization of building performance to support energy efficiency and carbon reduction goals.

The BEM process typically unfolds through a series of interconnected tasks that translate design intent and documentation into a simulation-ready input data format. It begins with information acquisition and preparation, where energy modelers gather project documents such as design briefs, audit reports, drawings, and specifications, and identify the relevant codes and reference data. The next stage is data structuring and extraction, which converts this heterogeneous and often unstructured information into structured simulation-ready data.

In current practice, however, few modeling workflows employ a unified "BEM schema" to store structured building information. Instead, modelers typically rely on customized spreadsheets or project-specific databases to organize simulation-ready inputs such as envelope parameters, internal loads, schedules, HVAC zoning and system details. These internal data tables are then manually or semi-automatically mapped into the input formats required by modeling software such as EnergyPlus/Openstudio, DesignBuilder, IES-VE, or eQuest. As a result, structured data exchange between tools remains limited, and interoperability challenges persist.

In parallel with structured data extraction, geometry definition is performed, where building geometry is reconstructed from drawings or BIM models and simplified into thermal zones suitable for simulation. Subsequent tasks include HVAC system modeling and baseline generation, which ensure consistency with applicable performance standards such as ASHRAE 90.1, followed by simulation and validation, where model behavior and results are checked for consistency and plausibility. Finally, post-processing and reporting synthesize simulation outputs into performance metrics, visualizations, and compliance documentation.

The concept of using a standardized schema to represent building data in a machine-readable and interoperable format is beginning to gain traction, particularly in the residential sector. One representative effort is the Home Performance eXtensible Markup Language (HPXML), developed by the Building Performance Institute (BPI), which defines a common data dictionary and transfer standard for home energy information (BPI, 2013). Such schema-based approaches provide a foundation for systematically transforming unstructured or semi-structured building documentation into structured representations, enabling more consistent data exchange and opening opportunities for automation across the BEM process.

Recent advances in large language models (LLMs) have opened new opportunities to automate many of these tasks. Their ability to interpret natural language, reason across heterogeneous data, and generate structured representations align closely with the information-rich and text-driven nature of BEM workflows. Zhang et al. (2024) developed an automated EnergyPlus input generation workflow that combined a language model with rule-based heuristics, though it lacked standardized metrics and reproducible datasets. Gang et al. (2024) further demonstrated how LLMs could support automatic metadata extraction and schema mapping for building energy models, highlighting their potential to reduce manual effort in model preparation.

As a result, LLMs have begun to emerge as promising tools for automating documentation parsing, model generation, and data validation. At the same time, research in natural language processing has developed more advanced prompting and reasoning strategies. Chain of thought prompting (Wei et al., 2022) and self-consistency (Wang et al., 2022) have both improved structured reasoning performance in general purpose tasks. Agentic workflow (Schick et al. 2023) also attracts more attention to conduct more complex tasks than prompting. These developments collectively lay the foundation for applying LLM reasoning capabilities to domain-specific modeling and simulation pipelines.

Although LLMs have been increasingly used in every field, including BEM, their accuracy and consistency are a persistent challenge due to hallucinations (Ji et al. 2023). This issue is particularly critical for engineering applications, which demand high levels of repeatability and precision. Quantitative evaluation of LLM performance through systematic benchmarking is therefore essential for assessing and improving their reliability. To evaluate LLM performance, there are many existing studies. In the AEC field, Arch-Eval (Wu et al., 2025) assessed architectural knowledge in large language models through domain-specific question answering. DrafterBench (Li et al., 2025) introduced a framework for interpreting civil engineering documents. Recent studies in related fields illustrate the growing interest in model evaluation.

While artificial intelligence has long been applied to energy prediction and optimization, the formal benchmarking of language models for BEM tasks has only recently begun. Current research remains fragmented, lacking a rigorous benchmark framework that integrates task-specific benchmark datasets, standardized performance metrics, and comprehensive coverage of end-to-end modeling workflows. Without such a framework, it is difficult to compare methods, quantify progress, or identify targeted opportunities for improvement.

To address this gap, we introduce BEMEval, an open benchmark framework designed to evaluate foundation models on building energy modeling tasks. It is designed to grow incrementally: each benchmark within the suite targets a distinct task with curated datasets, ground truth references, and quantitative metrics. As foundation model capabilities expand beyond text to include vision, multimodal reasoning, and physical simulation, BEMEval's scope will expand accordingly. It fills the research gap of lacking standardized benchmarking datasets and evaluation protocols, like ImageNet in the computer vision field (Deng et al. 2009), which ensures interoperability and reproducibility across tools and modeling platforms.

As the first step in this suite, we identify the most fundamental and time-intensive task: the mapping from unstructured building descriptions into structured building information, as the initial focus. This task directly leverages the core strengths of language models: contextual understanding, semantic extraction, and structured data generation.

In this work, we introduce BEMEval-Doc2Schema, the inaugural benchmark within the BEMEval framework, which evaluates accuracy of LLM models and prompting workflows in converting raw residential building documentation into standardized, machine-readable inputs for energy modeling. Residential buildings provide an ideal starting point for developing and evaluating LLM-based BEM automation. Compared with commercial or industrial facilities, residential buildings exhibit a higher degree of standardization in geometry, envelope assemblies, and HVAC system typologies. This consistency reduces modeling variability and allows more direct comparison between automated outputs and reference data. Moreover, the residential domain benefits from a relatively mature schema infrastructure and a growing ecosystem of open datasets. Standards such as HPXML (Home Performance eXtensible Markup Language) (BPI 2013) provide a rich hierarchical representation of home features, enabling consistent data exchange across energy audit, retrofit, and simulation tools. In addition, several well-documented residential testing facilities, including the HERS testing case L100, the NIST Net Zero Energy Residential Test Facility (NZERTF), and the NREL iUnit, offer detailed textual and quantitative building descriptions together with partially structured simulation-ready data.

In a practical BEM workflow, BEMEval-Doc2Schema corresponds to the early project phase in which energy modelers extract building characteristics from design documents, audits, and specifications and populate structured input templates or databases prior to simulation. Rather than replacing simulation engines or modeling expertise, LLM-based structured data extraction is intended to reduce manual transcription effort, improve consistency, and accelerate model setup. BEMEval provides a standardized way to evaluate whether such tools meet minimum engineering expectations before integration into production workflows.

The paper is organized as follows. Methodology introduces the BEMEval framework and scope and its first member BEMEval-Doc2Schema, including the structure of the benchmark and its key components. The Case Study section presents the evaluation of specific language models and prompting strategies using BEMEval-Doc2Schema. The Discussion section summarizes the main results and insights from the case studies. Finally, the Conclusion section highlights the primary contributions, demonstrates the utility of BEMEval, and outlines a long-term vision for developing a robust evaluation ecosystem for automated building energy modeling workflows.

# Methodology

## BEMEval Framework and Scope

BEMEval is a modular benchmark framework designed to evaluate foundation model performance across distinct building energy modeling tasks. Rather than treating BEM automation as a single monolithic problem, BEMEval decomposes the modeling workflow into task specific benchmarks, each with a clearly defined scope, curated datasets, and quantitative evaluation metrics. This modular structure enables controlled assessment of model capabilities at different stages of the BEM process while supporting incremental expansion of the benchmark suite.

The BEMEval framework is intended to grow over time to cover additional building energy modeling tasks that meet the same core selection criteria: high practitioner impact, benchmark plausibility, and suitability for foundation model evaluation. Two near-term candidates have been identified. The first is code compliance reasoning, which involves generating code compliant baseline configurations from building descriptions using standards such as ASHRAE 90.1 Appendix G. This task is well suited for benchmarking because the governing rules are publicly documented, deterministic, and text based, with ground truth derivable directly from the standard. The second is quality assurance and error detection, which focuses on reviewing simulation outputs to identify anomalies, unreasonable results, and modeling errors. This task is fundamentally reasoning driven, and benchmark ground truth can be constructed from expert annotated cases or intentionally seeded error models. Figure 1 illustrates the BEMeval framework.

As foundation model capabilities expand to include vision and multimodal reasoning, BEMEval may also target tasks involving architectural drawing interpretation, spatial reasoning, and geometry abstraction. Developing these future benchmarks will require close collaboration with practicing energy modelers to curate representative test cases and to validate evaluation metrics against expert judgment.

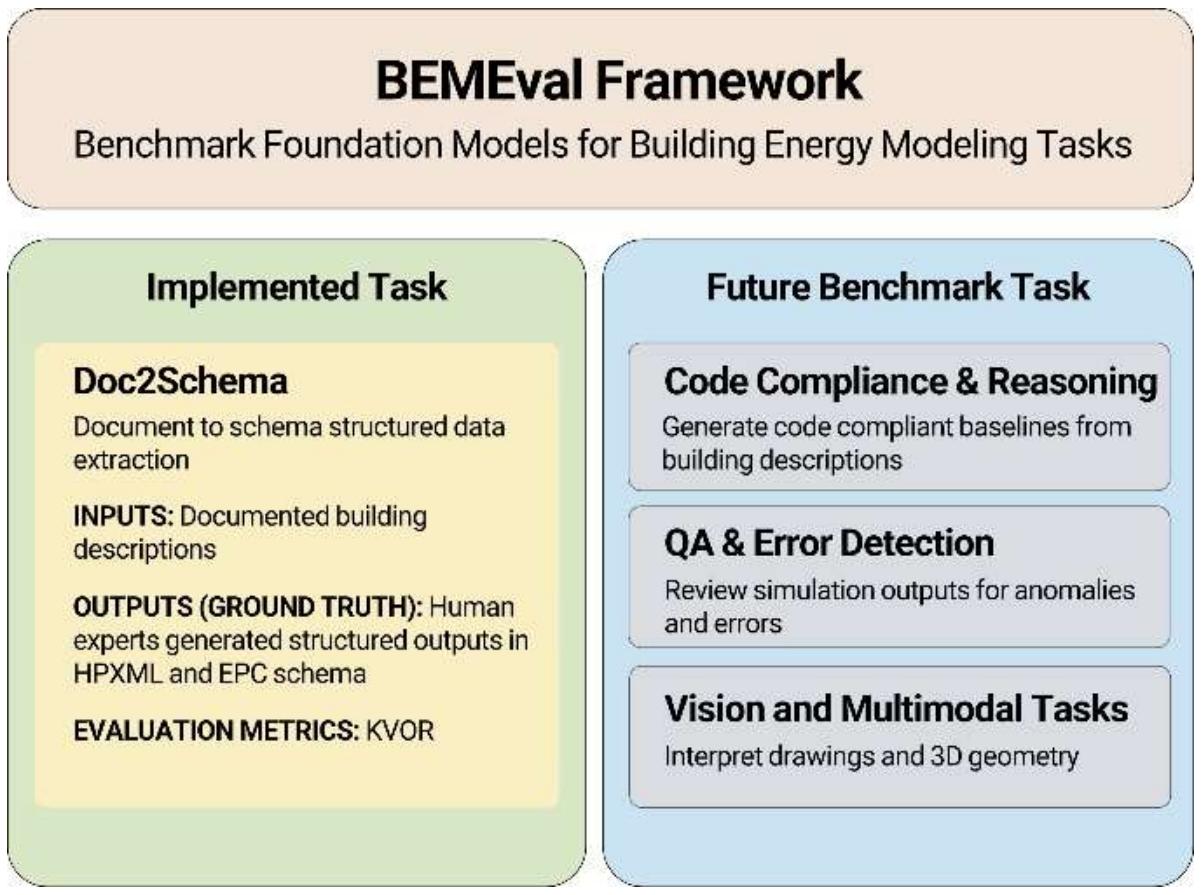

*Figure 1 BEMEval Framework*

BEMEval-Doc2Schema is the first benchmark implemented within the BEMEval framework. The selection of this task was guided by three criteria. First, it has high practitioner impact, as structured data extraction is one of the most labor intensive and repetitive tasks in current building energy modeling practice. Second, the task is benchmark plausible due to the availability of well documented building cases and standardized schemas that provide clear and auditable ground truth. Third, the task is well suited for large language models, as it primarily relies on text understanding, semantic interpretation, and domain reasoning.

The objective of the BEMEval-Doc2Schema benchmark is to provide a standardized test and grading framework for evaluating how effectively large language models and prompting regimes can convert unstructured building energy modeling documentation into structured and schema aligned data. The benchmark evaluates a model's ability to interpret building specific information, apply building energy modeling knowledge, and correctly populate target schemas without introducing unsupported assumptions.

BEMEval-Doc2Schema consists of three core components: inputs, reference outputs or ground truth, and evaluation metrics. The following sections describe each component in detail as shown in Figure 1.

**Inputs**: Real buildings' (L100 (PSD Consulting 2014), NZERTF (NIST 2014), and iUnit (DOE 2024)) unstructured BEM documentation in the formats of PDF, spreadsheet, etc.

**Outputs (Ground Truth):** The resulting structured datasets in the formats of existing widely used open schemas (HPXML and/or EPC). Components 1 and 2 are considered as input-output or question-answer pairs.

**Evaluation Metrics**: Definition of evaluation metrics to quantify how close the correct answer from Component 2 is from the LLM generated output.

## Benchmark Building Descriptions

### NZERTF

The NIST Net-Zero Energy Residential Test Facility (NZERTF) is a highly instrumented, single-family-style house located on the NIST campus in Gaithersburg, Maryland. Designed to resemble a typical suburban two-story home (approx. 2,700 ft² / 252 m² of living space) with basement and attic (each ~1,450 ft² / 135 m²) for a total volume of 44,900 ft³ (1,270 m³). The facility was conceived as a "laboratory home" to demonstrate and investigate whether a residential building of familiar size and form can produce as much energy through on-site renewables as it consumes over the course of a year (i.e., net-zero energy). It also serves as a research platform to validate building energy models, assess high-performance envelopes and system technologies, monitor indoor environmental quality, and explore the interaction of building systems including HVAC, ventilation, and photovoltaics.

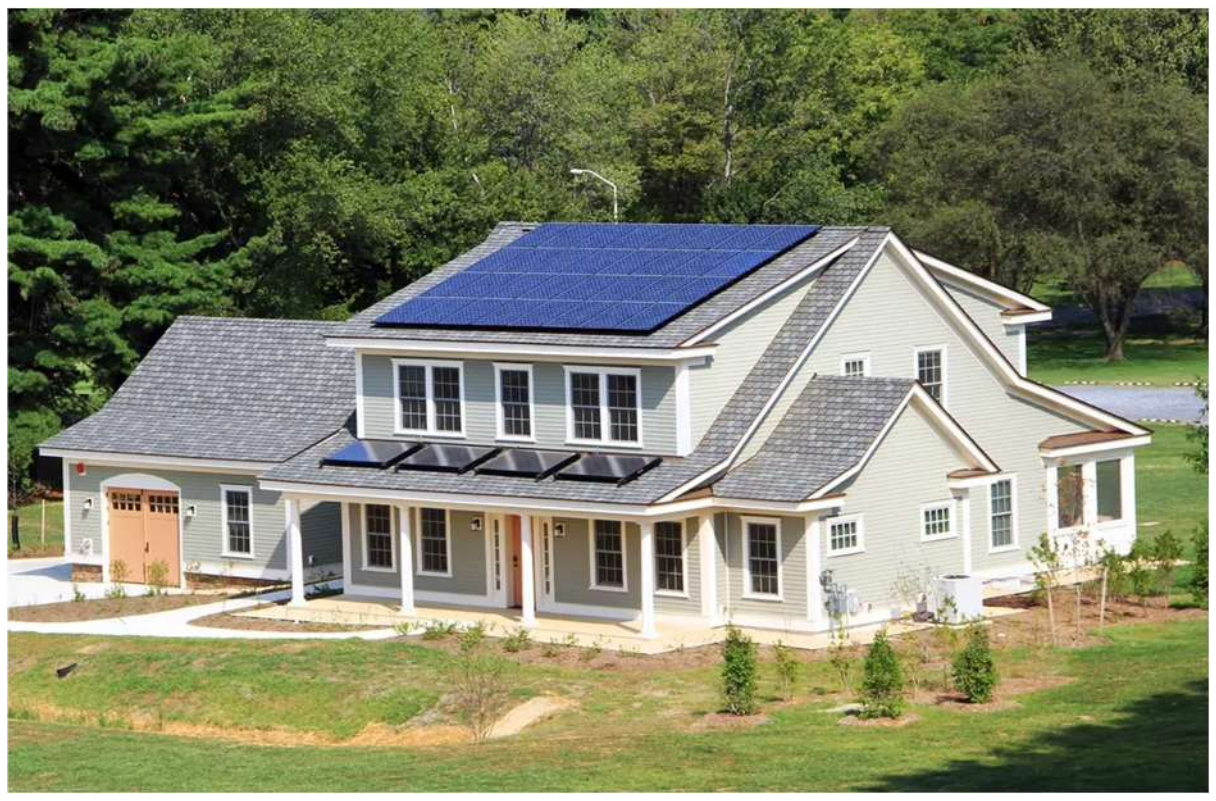

*Figure 2 NIST Net-Zero Energy Residential Test Facility*

Within the BEMEval-Doc2Schema framework, the NZERTF serves as a representative data point for the single-family residential typology. It provides a rich source of well-documented design and construction data, and also includes detailed equipment documentation for high-efficiency HVAC systems and renewable energy systems

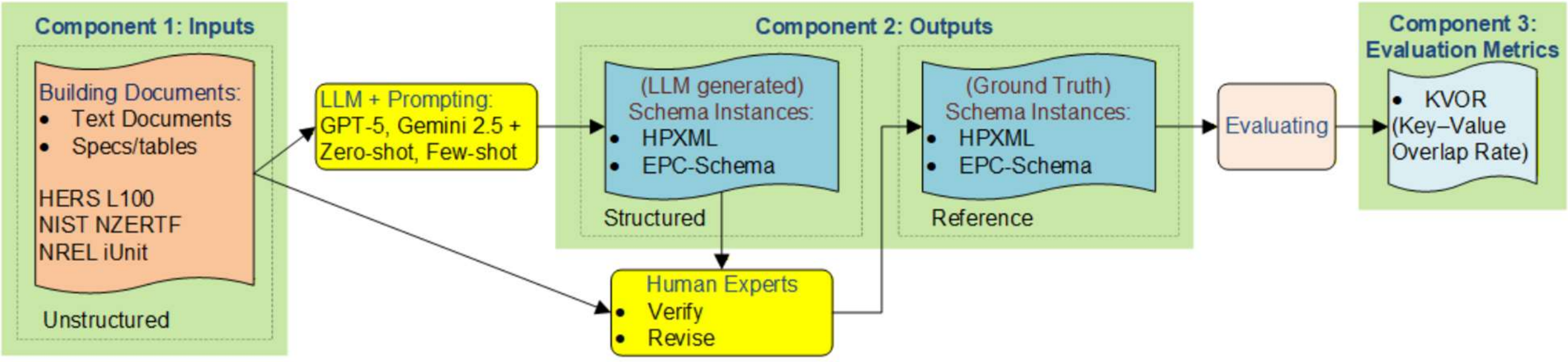

*Figure 2 BEMEval-Doc2Schema*

### iUnit

The NREL iUnit is a modular, factory-built apartment developed through a collaboration between the National Renewable Energy Laboratory (NREL) and the Denver-based developer iUnit. Designed as a high-performance, grid-interactive prototype, the iUnit represents a typical micro-apartment configuration within the multifamily housing sector. The prototype, approximately 38 m² (~400 ft²) in floor area, was constructed off-site and installed at NREL's Outdoor Research Block in Golden, Colorado.

The iUnit integrates advanced envelope design, efficient mechanical systems, and on-site renewable technologies to explore pathways for low-energy, factory-produced housing. Within the BEMEval-Doc2Schema framework, the iUnit serves as a representative data sample for the apartment and multifamily building typology.

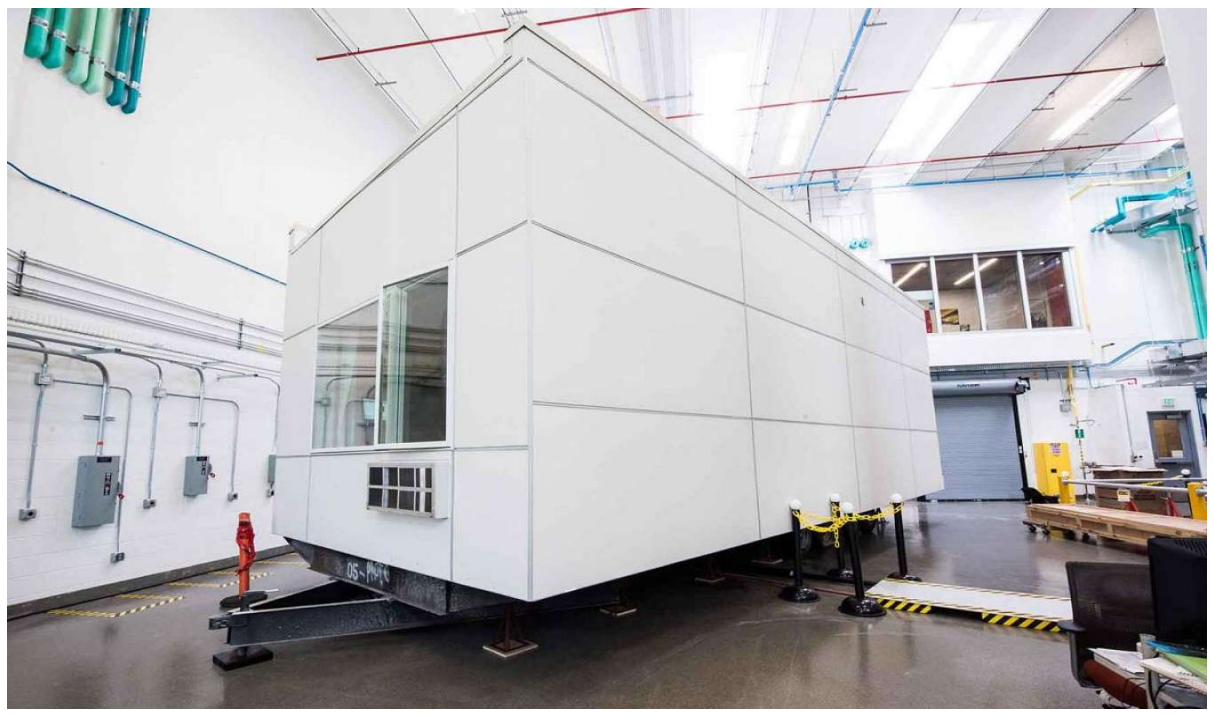

*Figure 3 NREL's iUnit modular apartment*

**L100**

The base building serves as the fundamental reference model for implementing the Home Energy Rating System (HERS) Building Energy Simulation Test (BESTEST), a method for evaluating the credibility of building energy software used by Home Energy Rating Systems.

The building represents a standardized residential structure used to evaluate and compare energy simulation results across different software tools. This is the case against which most other cases are compared to determine if the rating tool can accurately determine energy differences due to changes in building configuration.

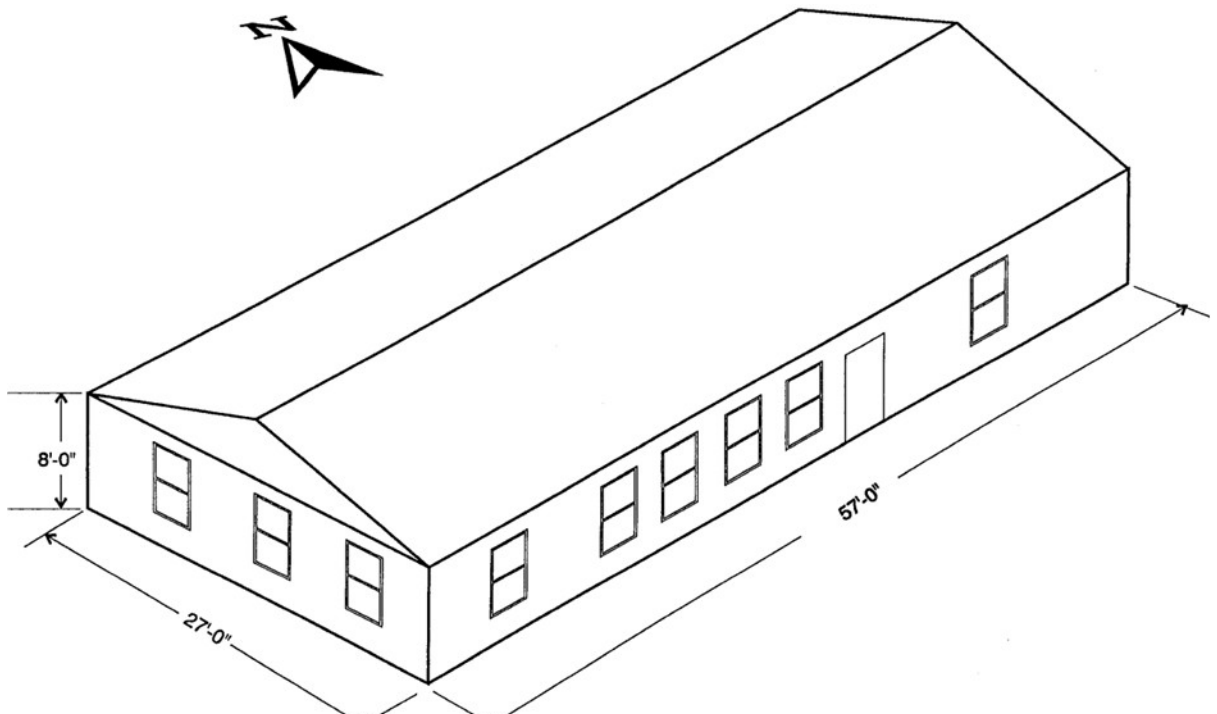

*Figure 4 NREL's base building L100 for HERS BESTTEST*

The base building, L100, is a 1539 ft², single-story, wood-frame, fully vented crawlspace home with 270 ft² of single-glazed windows (distributed with 90 ft² on the north and south faces and 45 ft² on the east and west faces). The walls are insulated with R-11 insulation, and the ceiling and floor are insulated with R-19 insulation.

It has one conditioned zone (the main floor), an unconditioned attic, and a vented crawl space. There are non-HVAC related internally generated loads of equipment, lights, people, animals, etc. For the base building, we consider only the pure load outputs of its mechanical systems, assuming all equipment, including ducts, is 100% efficient. Three additional assumptions are made: 1) the air system is fully (100%) convective; 2) the thermostat responds solely to air temperature; and 3) the thermostat is of the nonproportional type. Two locations are supplied for the test: Colorado Springs, Colorado (a clear and cold climate) and Las Vegas, Nevada (a hot and dry climate)

*Table 1 Summary of Schemas*

| | Description | Creator | Characteristics | Hierarchical? (Yes/No) | Customized? (Yes/No) | Links to Website |
|---|---|---|---|---|---|---|
| HPXML | HPXML (Home Performance XML) is a data transfer standard for home energy auditing and retrofit programs, designed to ensure interoperability among software tools, contractors, and program administrators in the residential energy efficiency sector. | BPI NREL DOE | Focuses on detailed residential audit, retrofit, and energy model data exchange; supports XML-based interoperability; aligns with RESNET and ENERGY STAR Home Upgrade programs. | Yes | No | https://hpxml.nrel.gov/ |
| EPC | EPC (Energy Performance Calculator) schema is a simulation-oriented data schema implementing normative energy calculation standards (CEN/ISO 13790, EN 15603, etc.) for evaluating building energy performance and certification across both residential and commercial buildings. | Georgia Tech High Performance Building Lab | Encodes parameters defined by ISO/CEN standards for thermal energy, delivered and primary energy, and emissions; supports simulation-based energy performance assessment. | No | Yes | https://github.com/BEMEval/doc2schema |

## Schema Descriptions

### HPXML

The Home Performance XML (HPXML**)** schema is an industry-standard data specification designed to facilitate information exchange among residential energy modeling, auditing, and rating tools. Developed through collaboration between the Building Performance Institute (BPI), the U.S. Department of Energy (DOE), and the National Renewable Energy Laboratory (NREL), HPXML provides a hierarchical, machine readable structure that describes residential building characteristics, including envelope assemblies, HVAC and service hot water systems, equipment efficiencies, fuel types, occupancy schedules, and other operational details, in a consistent and interoperable format. In the context of BEMEval-Doc2Schema, HPXML serves as the reference schema for evaluating the ability of language models to map unstructured residential building descriptions into structured, simulation-ready data. Because it encodes the same physical and system attributes that underpin detailed EnergyPlus or OpenStudio models, it provides a realistic and standardized representation of the residential modeling domain. Including HPXML in BEMEval-Doc2Schema therefore grounds the benchmark in an industry-recognized, widely adopted framework, ensuring that the evaluation tasks are aligned with current practice and immediately relevant to downstream production workflows.

### EPC-Schema

The Energy Performance Calculator (EPC) is a reduced-order building energy modeling framework developed by former Georgia Tech graduate students under the direction of Professor Godfried Augenbroe. The spreadsheet-based tool implements the CEN/ISO normative calculation standards (ISO 13790, EN 15603, etc.) for assessing building energy performance (Lee et al., 2011). Instead of replicating the detailed simulation approach used in conventional models, the EPC adopts a normative calculation paradigm that follows a systematic quasi steady state method to estimate energy performance based on standardized assumptions and simplified physical relationships. This approach bridges the gap between high-fidelity simulations and empirical benchmarking, providing a fast, transparent, and reproducible means to characterize building performance (Lee et al., 2013).

The EPC schema standardizes the inputs to an EPC simulation engine using TOML (Tom's Obvious, Minimal Language) as its data serialization format (TOML, 2021). TOML is a lightweight, human-readable configuration format that balances simplicity and strict typing, making it well suited for representing the structured input parameters required by the normative calculation method. Each EPC input file structures model parameters such as building envelope characteristics, system efficiencies, schedules, and climate data into a flat key value format. This format supports both interpretability and automation, ensuring that EPC models can be easily parameterized, version-controlled, and programmatically parsed.

The building descriptions, documentation, schema definitions, and evaluation resources are publicly available in the open-source BEMEval-Doc2Schema repository at https://github.com/BEMEval/doc2schema. The repository includes representative residential building cases (e.g., HERS L100, NIST NZERTF, and NREL iUnit), corresponding structured reference outputs in HPXML and EPC-Schema formats, schema specifications and metadata, and automated evaluation scripts implementing the Key–Value Overlap Rate (KVOR) metric. Comprehensive documentation and a getting-started tutorial are also provided to support reproducibility, extensibility, and community adoption.

### Comments

The reason for selecting HPXML and EPC lies in the fact that they are different in terms of level of details and focus. HPXML emphasizes detailed residential retrofit and audit-data exchange, enabling interoperability among contractors, program administrators, and software tools (BPI 2013, HPXML Guide 2023). On the other hand, EPC focuses on normative calculation methods and building energy performance rating across commercial and residential sectors.

Similarly, the three building cases were selected to represent different residential building types included in this study. This diversity ensures that the benchmark covers a representative range of residential configurations and data structures.

In future work, both the building portfolio and schema coverage will be expanded to include multifamily buildings, manufactured homes, and additional open schema formats.

**Evaluation Metrics**

The evaluation metric applied in this study is the Key–Value Overlap Rate (KVOR), which quantifies how closely the LLM–generated structured output matches the ground-truth reference. Because BEMEval-Doc2Schema focuses on mapping unstructured building documentation into schema-compliant formats such as HPXML and EPC-Schema, this metric provides a direct and interpretable measure of the model's extraction accuracy.

$$KVOR = \frac{Number\ of\ correctly\ matched\ key-value\ pairs}{Total\ number\ of\ keys\ in\ the\ reference\ sche} \quad (1)$$

In this equation, the numerator represents the number of key–value pairs in which both the key and its corresponding value are correctly identified by the model and align with the ground truth. The denominator represents the total number of keys defined in the reference schema for a given dataset. The resulting ratio ranges from 0 to 1, with higher values indicating closer agreement between the LLM-generated output and the expected structured data.

Even when an LLM output passes schema alignment validation, it may still assign a value from the unstructured document to an incorrect field due to semantic misunderstanding, contextual ambiguity, or domain knowledge limitations. The Key–Value Overlap Rate captures such discrepancies by penalizing incorrect associations between keys and values, offering a quantitative measure of both structural and semantic accuracy.

A higher Key–Value Overlap Rate reflects the model's stronger capability to interpret building information and correctly map it to the designated schema fields. Conversely, a lower rate indicates that the model struggles with maintaining consistent correspondence between document content and schema structure. This metric thus provides a clear and reproducible basis for comparing different LLMs and prompting strategies in automated building energy modeling workflows.

## Case Study

We conducted evaluations of two LLM models and two prompting workflows under our developed BEMEval to test their capability to convert L100, iUnit, and NZERTF's raw building description to the structured schemas of HPXML and EPC in this section.

The two foundation models we use are GPT-5 and Gemini 2.5. GPT-5, developed by OpenAI and released in 2025, extends the GPT series with enhanced reasoning, longer context handling, and improved structured data generation, making it effective for tasks involving schema alignment and technical text interpretation. Gemini 2.5, introduced by Google DeepMind in 2025, integrates multimodal understanding with symbolic reasoning and factual grounding. It is designed for high consistency in analytical reasoning and excels in translating complex documentation into structured, machine-readable formats.

The two prompting strategies are: 1) zero-shot prompting, 2) few-shot prompting. Zero-shot prompting means the model receives only task instructions and schema documentation, without prior examples. It relies purely on internal knowledge and reasoning to complete the mapping. Few shot prompting means that the model is additionally provided with a set of input–output examples demonstrating correct schema formatting and mapping behavior. These examples guide the model to learn mapping conventions before handling new input.

The zero-shot prompting template is shown in the listing below:

*System / Role Instruction:*

You are a domain expert in building energy modeling, data standardization, and {schema_name(e.g., HPXML)}schema

*Task:*

Your task is to convert unstructured building energy modeling information — which may come from PDFs, Excel files, Word documents, plain text, or scanned descriptions — into a structured {schema_name(e.g., HPXML) format in an {file_format(e.g., .xml)} file.

The unstructured building information is uploaded as an attachment called {name_of_unstructured_data.pdf}

The HPXML schema is described in the attachment called: {schema_documentation.pdf}

*Follow the following rules strictly:*

1. Extraction and Mapping Rules

- The output should strictly follow the HPXML format

- Do not assume or infer any value that is not explicitly stated in the input
- Ignore building geometry-related input such as coordinates

2. Every HPXML field must be traceable — include an XML comment (<! -- -->) above each field that explains:

- The exact source text or cell where the value was found, and
- The reasoning for placing it in that specific HPXML element.
- If a value is missing, omit that field entirely (do not fill defaults)

*Listing 1. Zero-Shot Prompting Template*

The additional prompts add to Listing 1 for few-shot prompting template is shown in the listing below:

Below are [N] example pairs of unstructured input and the corresponding HPXML output for reference.

The model should learn the format, comment style, and mapping behavior from these examples before handling new input.

Example input 1: {Example_input_1}

Example output 1: {Example_output_1}

…

Example input n: {Example_input_n}

Example output n: {Example_output_n}

*Listing 2. Few-Shot Prompting Template*

**Results**

*Table 2. Summary of Model Performance (Key–Value Overlap Rate, KVOR)*

| **Score: KVOR** | **L100 (HPXML)** | **iUnit (EPC)** | **NZERTF (HPXML)** |
|---|---|---|---|
| Zero Shot - ChatGPT | 0.000 | 0.480 | 0.005 |
| Zero Shot - Gemini | 0.040 | 0.851 | 0.166 |
| Few Shot - ChatGPT | 0.010 | NA | NA |
| Few Shot - Gemini | 0.082 | NA | NA |

The results demonstrate clear differences in both model capability and prompting strategy:

**Model Comparison**: Across all datasets, Gemini 2.5 consistently outperformed GPT-5, demonstrating stronger schema comprehension and more consistent value mapping. Gemini's notably high score on the iUnit EPC dataset (KVOR = 0.851) suggests that it is better able to interpret structured key value relationships and maintain contextual consistency within the flat EPC schema format.

**Prompting Strategy**: Both models showed improved accuracy under few-shot prompting, confirming the effectiveness of example-based guidance for schema alignment. The improvement was especially noticeable in Gemini's L100 case (from 0.040 to 0.082).

**Schema Complexity**: Performance was generally higher for the EPC schema than for HPXML, primarily due to EPC's simpler and more normalized structure. HPXML's deeper hierarchy, greater attribute diversity, and embedded domain conventions posed more challenges for the models, leading to lower overlap rates.

**Building Case Difficulty**: Among the HPXML-based datasets, NZERTF yielded the lowest overlaps, reflecting its greater system complexity and higher density of interdependent parameters (e.g., HVAC, ventilation, renewable systems).

Overall, the results highlight that LLM-based schema extraction remains sensitive to data structure, schema depth, and prompt design. While Gemini shows promising generalization and accuracy, consistent improvement will require domain-informed fine-tuning, unit normalization, and hybrid validation pipelines integrated into future versions of BEMEval.

## Discussion

The results of this study highlight both the potential and current limitations of using large language models to automate the translation of unstructured building descriptions into standardized energy modeling schemas.

Across all cases, Gemini 2.5 outperformed GPT-5, demonstrating stronger schema comprehension and contextual reasoning. Its notably high KVOR on the iUnit EPC dataset (0.851) suggests better adaptation to simple and flat data structures. HPXML's broader hierarchy and cross-referenced identifiers across systems introduced greater relational complexity, making consistent mapping more difficult. Among HPXML cases, NZERTF produced the lowest scores, reflecting its detailed systems and highly parameterized structure.

Both models benefited from few-shot prompting, confirming that example-based guidance improves schema alignment. However, the limited availability of examples constrained evaluation to a single few-shot configuration, offering only a partial view of the models' in-context learning ability.

To interpret these results appropriately, it is important to note the scope and limitations of the KVOR metric. KVOR treats all schema keys equally and does not weight parameters by their relative impact on simulation outcomes. While this simplifies interpretation and enables schema agnostic comparison, future work will explore sensitivity aware weighting schemes, parameter grouping, and downstream simulation impact metrics. The current implementation is intentionally conservative and task focused, prioritizing extraction correctness over performance attribution.

Although the primary goal of this study is to establish a benchmarking framework rather than a fully automated system, the findings provide valuable insight into how future workflows and data schemas might evolve to better accommodate AI-based automation tasks. Across different model configurations, including zero shot, few shot prompting workflows, a consistent trend appears: models perform more effectively when working with schemas that have higher abstraction and lower structural complexity. These results suggest that, until language models become more capable of handling complex data structures, greater attention should be given to designing "AI-friendly" schemas that can better support partial automation. Simplified, semantically clear, and well-documented schemas could potentially help bridge the current gap between natural-language inputs and structured model representations. While schema design was not the central focus of this study, it emerged as a potential area for future exploration. A possible direction is establishing schema evaluation metrics that measure how different schema design features contribute to improved automation performance.

Another noteworthy observation is that large language models tend to favor automation over manual reasoning when performing structured data generation tasks. When asked to convert unstructured building descriptions into schema-based representations, the models frequently resorted to writing code or scripts that automate data extraction and organization rather than directly carrying out the mapping themselves. This behavior suggests that the models interpret such tasks as a program synthesis problem, framing them as opportunities to generate procedures that could accomplish the task rather than engaging in explicit semantic reasoning. While this reflects an effort to find a resource efficient or computationally economical way to perform the task, it also reveals a form of bias or shortcut thinking, where the model assumes that the problem can be addressed through code rather than understanding the underlying domain semantics.

A central design principle of BEMEval-Doc2Schema is reproducibility and practical integration into real-world BEM workflows. All benchmark datasets are derived from well-documented residential test cases with clearly defined inputs, schema references, and evaluation criteria, enabling consistent comparison across models, prompting regimes, and experimental settings. The benchmark aligns with common BEM practices by explicitly targeting intermediate, simulation-ready data representations—such as HPXML and EPC-Schema—that can be directly consumed by downstream tools including EnergyPlus or OpenStudio-based workflows. By focusing on schema-level correctness rather than tool-specific input files, BEMEval-Doc2Schema decouples language model evaluation from individual simulation engines while preserving compatibility with existing modeling pipelines. This design enables BEMEval-Doc2Schema to serve both as a controlled research benchmark and as a validation layer within automated or semi-automated BEM production workflows.

BEMEval-Doc2Schema is released as an open-source, community-oriented framework to encourage transparent evaluation, extensibility, and broad adoption. The benchmark resources, including datasets, schema definitions, evaluation scripts, and documentation, are publicly available to support independent reproduction of results and facilitate the contribution of new tasks, schemas, and building cases. By adopting open standards and modular design, BEMEval-Doc2Schema enables researchers and practitioners to extend the benchmark to additional building types, modeling stages, and language model architectures. Over time, this open benchmarking ecosystem is intended to support cumulative progress in AI-assisted building energy modeling, enabling the community to track performance improvements, identify persistent failure modes, and establish shared baselines for future methodological advances.

## Limitations and Future Work

While this study presents an initial effort toward evaluating the use of large language models in building energy modeling, several limitations should be acknowledged.

The current version of the BEMEval-Doc2Schema benchmark includes a limited number of data points. While these examples capture representative building types and schemas, they do not yet reflect the full diversity of residential building energy modeling practice. For example, the HERS L100 case serves as a standardized reference but does not include explicit HVAC system information. The NIST NZERTF case provides detailed envelope and system data but represents a single, highly instrumented prototype dwelling optimized for a specific climate, which limits its generalizability to broader residential contexts, and the NREL iUnit case features simplified internal loads and schedules developed primarily for calibration studies. Future efforts will focus on expanding the dataset to include a wider range of building types, vintages, and system configurations that will improve both representativeness and generalizability.

The limited number of data points also constrains the evaluation approach. In the few-shot setup, the availability of examples was further restricted: only one test configuration was performed, using HERS L100A as the target case and the NZERTF input–output pair as few-shot examples. The reverse configuration (NZERTF as target with L100A examples) was not tested, and the iUnit case, which employs the EPC-schema, lacked corresponding examples for few-shot evaluation. Consequently, the few-shot results represent only a partial view of the models' in-context learning potential.

The models evaluated in this study were general-purpose LLMs without any domain specific fine-tuning. While they demonstrate broad linguistic and reasoning abilities, they lack specialized understanding of building physics, HVAC systems, and schema conventions. Their performance therefore reflects baseline generalization rather than the upper bound of what could be achieved through targeted adaptation. Future work should explore models that are fine-tuned or instruction-optimized using building design and simulation corpora, as well as schema documentation to better align with the semantics of HPXML and EPC style datasets. Incorporating such strategies, together with expanded datasets and more balanced few shot configurations, will enable a deeper evaluation of how domain adapted language models can improve accuracy, consistency, and interpretability in automating building energy modeling workflows.

To quantitatively evaluate the accuracy of the extracted data, we define the evaluation rate as the ratio between the number of the correct key values in the LLM answer and the number of expected key values of the schema. The definition of a correct value depends on the data type. For string-based attributes, some string similarity metrics are employed to assess textual correspondence, including edit-distance–based measures such as the Levenshtein distance (Levenshtein 1966) and Jaro–Winkler similarity (Winkler 1990), token-based measures such as the Jaccard and Sørensen–Dice coefficients (Jaccard 1912, Sørensen 1948), and vector-based measures such as cosine similarity (Salton et al. 1983). These metrics are implemented using Python libraries such as RapidFuzz and TextDistance (RapidFuzz Documentation 2025; TextDistance Documentation 2025), which provide efficient and reproducible similarity computations. A similarity threshold (e.g., $\geq$0.9 for normalized string metrics or $\leq$5% for numerical deviations) is applied to determine whether an extracted value is considered correct. In future work, an automated validation toolchain will be developed to integrate these similarity metrics, unit normalization procedures, and configurable thresholds to systematically verify the result correctness.

However, these traditional similarity metrics primarily capture surface-level lexical or syntactic correspondence, often failing to account for semantic equivalence or contextual meaning in building-related text (Reimers et al. 2019). For instance, expressions such as "mechanical ventilation" and "HVAC system" may refer to similar concepts but yield low string similarity scores. In addition, numerical comparisons may overlook unit conversion inconsistencies or context-dependent tolerances, which are common in engineering datasets.

Future improvements may include incorporating embedding-based semantic similarity measures using transformer encoders (e.g., Sentence-BERT) and domain-specific normalization rules to better reflect true equivalence between extracted and reference values (Zhang et al. 2020). In addition, future extensions of BEMEval will introduce robustness-oriented evaluation dimensions, such as multi-seed testing to assess output stability, systematic measurement of hallucination rates, and analysis of cost–latency–accuracy tradeoffs across models. An automated validation toolchain will be developed to integrate these similarity metrics, unit normalization procedures, robustness indicators, and configurable thresholds to systematically verify result correctness under realistic deployment conditions.

Beyond expanding its technical scope, the long-term success of the benchmark will depend fundamentally on sustained community collaboration. Cultivating an active, inclusive, and well supported open-source community is essential to ensure that the benchmark continues to evolve in relevance and rigor. Comprehensive documentation, accessible tutorials, and open access to evaluation resources can lower barriers to participation and empower researchers, practitioners, and developers to contribute, validate, and extend the benchmark collectively.

## Conclusion

The main contribution of this work is the proposal and initial implementation of BEMEval, a framework and dataset collection for evaluating large language models in building energy modeling tasks. Through consistent evaluation protocols, transparent metrics, and reproducible datasets, BEMEval establishes a standardized foundation for comparing model performance across a range of BEM tasks.

The benchmark's first implementation, BEMEval-Doc2Schema, applies the framework to evaluate how language models perform in mapping residential building data into structured schema representations. Results indicate that models can complete portions of the task when the schema is relatively simple and example guidance is provided, but their performance declines as schema complexity increases. These findings highlight the importance of systematic benchmarking for revealing current limitations and guiding the development of more robust and reliable AI assisted modeling workflows.

BEMEval has the potential to serve as a bridge between artificial intelligence and building simulation communities. By framing modeling tasks as measurable and reproducible challenges, it encourages collective participation and fosters a culture of transparency and collaboration. The benchmark aims to provide a foundation for sustained engagement among researchers, practitioners, and software developers, enabling progress to be monitored as modeling standards and AI technologies continue to evolve.

Looking ahead, the long-term goal of this work is to contribute to a robust and extensible open-source evaluation ecosystem for building energy modeling. As foundation models and domain datasets advance, BEMEval can evolve into a community resource for tracking progress and enhancing the integration of artificial intelligence into building modeling practice, enabling more efficient, transparent, and accessible workflows in the future.